\documentclass[twocolumn,aps,prl,groupedaddress,showpacs]{revtex4}
\usepackage{epsfig,amssymb,amsmath}
\usepackage{amsmath}
\usepackage{graphicx}
\usepackage{amssymb}

\usepackage{epsfig}
\usepackage{latexsym}
\usepackage{amssymb}
\usepackage{graphicx}

\usepackage{color}
\usepackage{ulem}

\begin{document}

\title{Controlled coupling of spin-resolved quantum Hall edge states}

\author{
Biswajit Karmakar$^{1}$, Davide Venturelli$^{1,2,3}$, Luca Chirolli$^{1}$, Fabio Taddei$^{1}$, Vittorio Giovannetti$^{1}$, Rosario Fazio$^{1}$, Stefano Roddaro$^{1}$, 
Giorgio Biasiol$^{4}$, Lucia Sorba$^{1}$, Vittorio Pellegrini$^{1}$, and Fabio Beltram$^{1}$
}
\affiliation{$^{1}$NEST, Scuola Normale Superiore and Istituto Nanoscienze-CNR,  Piazza San Silvestro 12, I-56127 Pisa, Italy, \\
$^{2}$Institut NEEL, CNRS and Universit\'{e} Joseph Fourier, Grenoble, France, \\
$^{3}$International School for Advanced Studies (SISSA), Via Bonomea 265, I-34136 Trieste, Italy, \\
$^{4}$Istituto Officina  dei Materiali CNR, Laboratorio TASC, Basovizza (TS), Italy.}

\date{\today}

\begin{abstract}
We introduce and experimentally demonstrate a new method that allows us to controllably couple co-propagating spin-resolved edge states of a two dimensional electron gas (2DEG) in the integer quantum Hall regime. The scheme exploits a spatially-periodic in-plane magnetic field that is created by an array of Cobalt nano-magnets placed at the boundary of the 2DEG. A maximum charge/spin transfer of $28\pm 1 \%$ is achieved at 250 mK.
\end{abstract}


\maketitle

Topologically-protected edge states are dissipationless conducting surface states immune to impurity scattering and geometrical defects that occur in electronic systems characterized by a bulk insulating gap~\cite{KANE}.
One example can be found in a clean two-dimensional electron gas (2DEG) under   high magnetic field in the quantum Hall (QH) regime~\cite{KLIT}.
In the integer QH case,
spin-resolved edge states (SRESs) at filling fraction $\nu = 2$ (number of filled energy levels in the bulk) 
 are characterized by very large relaxation~\cite{MullerPRB1992} and coherence~\cite{Heiblum2003} lengths. 
This  system  is a promising building block for the design of coherent electronics circuitry~\cite{Heiblum2003,ROU,LIV,roddaro,BIE}. 
 It represents also  an ideal candidate for the implementation of dual-rail quantum-computation architectures~\cite{DUALRAIL} by encoding 
 the qubit  in the spin degree of freedom that labels two distinct co-propagating, energy-degenerate SRESs
 of the {\it same} Landau level (LL) at the {\it same} physical edge of the 2DEG~\cite{VG}.
\begin{figure}[t]
 \begin{center}
  \includegraphics[width=7cm]{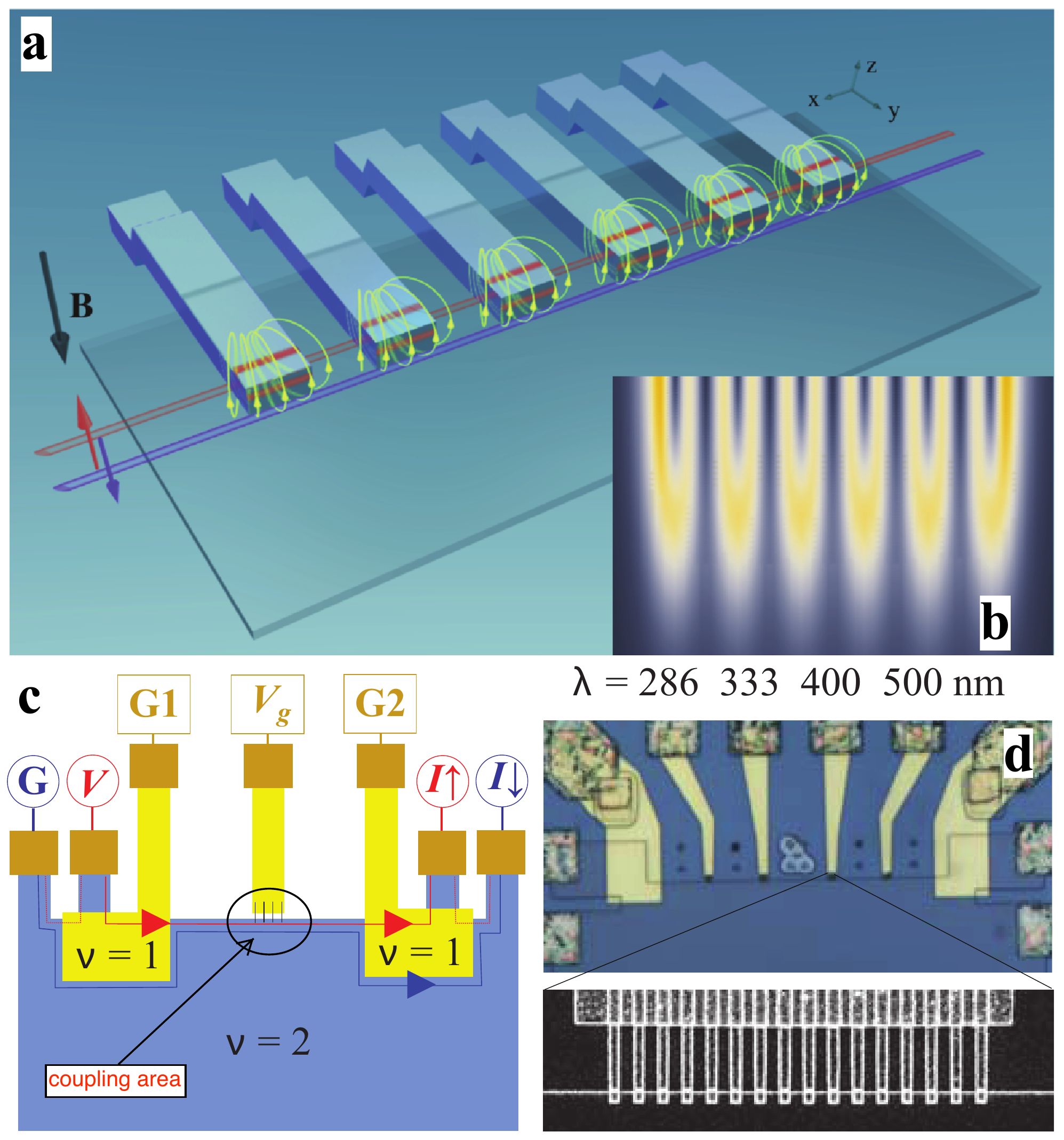}
    \caption{(Color online) a) Schematics   of the device. 
    The Cobalt fingers (blue bars) produce a fringing  field (yellow lines) resulting in an in-plane, oscillatory, magnetic field $\vec{B}_{\parallel}$ at the level of the 2DEG (textured gray) 
    residing below the top surface.
    The field induces charge transfer between the
      spin up 
      $\Psi_{\uparrow}$  SRES
     (red line) and
      spin down 
       $\Psi_{\downarrow}$ SRES
      (blue line). b) Density plot of the modulus $\vec{B}_{\parallel}$ in the proximity of the magnetic fingers on 2DEG  plane. The dashed line indicates the end of the finger array
      at $0.2$ $\mu$m from the physical edge of the mesa (white stripe). c) Measurement set-up: 
      The  $\Psi_{\uparrow}$ channel
       is excited by a bias voltage $V$, while  $\Psi_{\downarrow}$ 
       is grounded at the contact denoted by G. The SRESs can be reversibly decoupled by negatively biasing the array with a voltage $V_a$
       (G1 and G2 are contacts for the top gates). d) Optical image of the device showing four sets of magnetic fingers  with different periodicity $\lambda$ placed serially at the mesa boundary
       (the yellow elements are  gold eletrical contacts). 
       Zoomed region is the scanning electron microscopic image of the
               array of periodicity $\lambda=400~{\rm nm}$: it
        is nearly $6 ~\mu{\rm m}$ long and has an overlap on the mesa of $0.2 ~\mu$m.} \label{fig1}
 \end{center}
\end{figure}
A key element for the realization of such architecture~\cite{STACE,nayak,VG} is a coherent beam splitter
   that makes it possible to prepare any superposition of the two logic states, thus realizing one-qubit gate transformations.
This requires the ability to induce controlled charge transfer between the two co-propagating SRESs,  a goal which up to date has not been yet
achieved. Here we  solve the problem by  targeting a resonant condition, in analogy with 
 the periodic poling technique adopted  in optics
 ~\cite{PERIOD}.

In the integer QH regime the SRESs are
single-particle eigenstates $\psi_{nks}(x,y)=|s\rangle \; e^{ikx} \; \chi_{nk}(y)/\sqrt{L}$ of the Hamiltonian
$H=({\bf p}+e{\bf A})^2/2m^*+V_c(y)-\frac{1}{2}g^*\mu_BB\sigma_z$ which describes a 
2DEG in the $(x,y)$-plane, subject to a strong magnetic field $B$ in the $z$-direction and confined  transversely by the  potential $V_c(y)$~\cite{QHE}. Here 
${\bf p}\equiv(p_x,p_y)$ and $\vec{\sigma}\equiv(\sigma_x,\sigma_y,\sigma_z)$ are respectively,  the particle momentum and spin operators, ${\bf A}$ is the vector potential, $L$ is the longitudinal  length of the Hall bar,
while  $m^*$ 
and $g^*$  are  the effective electron mass and g-factor of the material.
Each $\psi_{nk s}(x,y)$ represents an electron state of the $n$th LL 
with spin projection $s\in\{\uparrow,\downarrow\}$ along $z$-axis, which is characterized by a transverse spatial distribution $\chi_{nk}(y)$, and which 
propagates along the sample
with longitudinal wave-vector $k$.
In our analysis we will focus on a $\nu=2$ configuration, where the longitudinal electron transport  occurs through the SRESs of the lowest LL,  i.e. 
$\Psi_{\uparrow} \equiv \psi_{0,k_{\uparrow},\uparrow}(x,y)$ and $\Psi_{\downarrow} \equiv \psi_{0,k_{\downarrow},\downarrow}(x,y)$
(the values  $k_{\uparrow}, k_{\downarrow}$  being 
determined  by  the degeneracy condition at the Fermi energy $E_F = \epsilon_{k_{\uparrow}}=\epsilon_{k_{\downarrow}}$ of the 
 corresponding eigenenergies). Specifically  in our scheme   
the two SRESs are separately contacted, grounding $\Psi_{\downarrow}$ and
injecting electrons on $\Psi_{\uparrow}$ via a small bias gate $V$.
      The spin resolved currents $I_{\uparrow}$ and $I_{\downarrow}$ of the two SRESs are then separately measured at the output of the device, 
          after an artificial charge transfer  from $\Psi_{\uparrow}$ to $\Psi_{\downarrow}$
           is induced during the propagation.  
Since in general $\Delta k\equiv k_{\uparrow}- k_{\downarrow}\neq0$, 
$\Psi_{\uparrow}$ and $\Psi_{\downarrow}$ support electrons 
at different wave vectors. 
Hence any  external perturbation 
capable of inducing charge transfer between them
must both flip the spin {\it and} provide a suitable momentum transfer to match the wave-vector gap  $\Delta k$.
In our scheme we achieve this by introducing 
a spatially-periodic in-plane magnetic fringing field ${\vec{B}}_\|(x,y)$~\cite{Cocriticalfield}
generated by  an array of Cobalt nano-magnet ({\it magnetic fingers}) placed along the longitudinal direction of the 2DEG,
see Fig.~\ref{fig1}a.
The system Hamiltonian acquires thus a local perturbation term
$\Delta H=-g^*\mu_B{\vec{B}_\|(x,y)} \cdot {\vec{\sigma}}/2$,
which at first order induces a transferred current $I_{\downarrow}=(e^2V/h)|t_{\uparrow\downarrow}|^2$,
where $t_{\uparrow\downarrow}=(L/i\hbar v)\langle\Psi_{\downarrow}|\Delta H|\Psi_{\uparrow}\rangle$ is the associated 
scattering amplitude, and 
$v$ is the group velocity of the SRESs. 
To capture the essence of the phenomenon, consider  for instance an array of periodicity $\lambda$ and longitudinal extension  $\Delta X$ described by a ${\vec{B}}_\|(x,y)$
 field of the form  
$B_{y}(y)\cos(2\pi x/\lambda) \hat{y}$ for $x\in [ -\Delta X/2, \Delta X/2]$ and zero otherwise (here for simplicity $x$ and $z$  component of $\vec{B}_\|$
 have been neglected). 
The corresponding transmission amplitude computed at lowest order in the T-matrix expansion~\cite{DIVENTRA}  is
\begin{equation}\label{Eq:t12amp}
t_{\uparrow\downarrow}= ig^*\mu_B\langle B_y\rangle\tfrac{\Delta X}{4\hbar v} \; \mbox{sinc}[(2\pi/\lambda -\Delta k){\Delta X/2}]\;,
\end{equation}
with $\mbox{sinc}[\cdot]\equiv \sin[\cdot]/[\cdot]$ being the sine cardinal function and 
$\langle B_y\rangle\equiv \int dy B_y(y)\chi_{0,k_{\uparrow}}(y)\chi_{0,k_{\downarrow}}(y)$.
The expression clearly shows that even for small values of longitudinal field a pronounced enhancement in inter-edge transfer occurs 
 when $\lambda$ matches the wave-vector difference of the two SRESs (i.e. $ \lambda _{res} = {2\pi}/{\Delta k}$),
the width of the resonance being inversely proportional to $\Delta X$.  

The quantity $\Delta k$ that defines the resonant condition depends on the Zeeman energy gap and on the details of the confinement potential $V_c(y)$. 
An estimate based on numerical simulations  (see Supplemental Material (SM))
leads to an approximate value $\lambda_{res}  \approx $ 400 nm at B = 4.5 T,
 which we assumed as a starting point in designing our setup.
The  device was fabricated on one-sided modulation-doped AlGaAs/GaAs heterostructure grown by molecular beam epitaxy. The 2DEG resides at the AlGaAs/GaAs heterointerface located 100 nm below the top surface. A spacer layer of 42 nm separates the 2DEG from the Si $\delta$-doping layer above it. The 2DEG has nominal electron density of $2\times10^{11}/{\rm cm}^2$ and low-temperature mobility nearly $4\times10^6~{\rm Vcm/s}$. The Cobalt nano-magnet array was defined at the mesa boundary of the 2DEG using e-beam lithography and thermal evaporation of 10 nm Ti followed by 110 nm Co.
Eight nano-magnet arrays at different periodicities 
(specifically $\lambda=
500, 400, 333, 286, 250, 222, 200$ and  $182~\text{nm}$)
were fabricated, keeping the total spatial extension of the modulation region nearly constant, $\Delta X \simeq  6.2~\mu$\rm{m} (four of them are on the other side of the mesa and therefore not visible in the microscope image of Fig. \ref{fig1} d).
The magnetization of the Cobalt fingers is aligned along the applied perpendicular magnetic field $B$ (Fig.~\ref{fig1}a), if $B$ is large enough~\cite{Cocriticalfield}. 
The actual value of the oscillatory $\vec{B}_{\parallel}$ can reach $50~\text{mT}$ in the proximity of the fingers and 
it decays away from the array (see Fig.~\ref{fig1}b). Importantly, 
coupling between the SRESs and a chosen set of fingers can be activated by increasing 
  the voltage bias $V_a$ of  the array from 
$-3$ to $0$ V (Fig.~\ref{fig1}c), while keeping all other arrays at $V_a=-3$ V.
In these conditions, the SRESs are brought close to the selected array only and exposed to its oscillatory in-plane field $\vec{B}_{\parallel}$.
Transport measurements were carried out in a He3 cryo-system with a base temperature of 250 mK equipped with 12 T superconducting magnet. An ac voltage excitation of $25.8~\mu{\rm V}$ at $17~{\rm Hz}$ was applied to the electrode V of Fig.~\ref{fig1} c) and the transmitted current was measured by standard lock-in techniques using  current to voltage preamplifiers.

\begin{figure}[t]
 \begin{center}
  \includegraphics[width=7 cm]{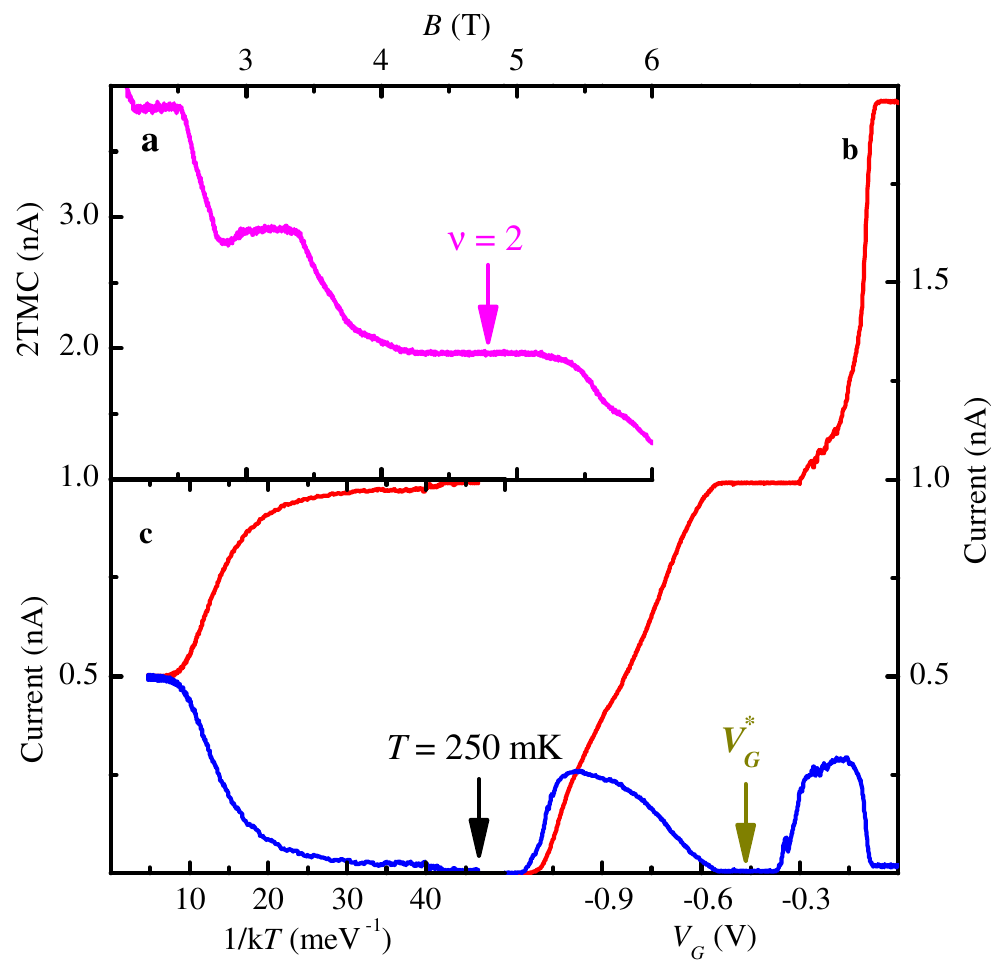}
    \caption{(Color online) a) Plot of the two terminal magneto-current (2TMC) measured at 250 mK.
    The value of magnetic field $B=4.75~{\rm T}$, indicated by an arrow, is used to place the 2DEG approximately at the center of the $\nu=2$ plateau. b) Plot of the currents $I_{\uparrow}$ (red) and $I_{\downarrow}$ (blue) measured at the current terminals red and blue respectively (Fig.~\ref{fig1}c) with the voltage $V_G$ applied to the gates G1 and G2, while the nano-magnets are deactivated by applying a voltage bias of $V_a = -3$ V to all the arrays.
The value of $V_G$  is set to $V^*_G$, indicated by an arrow, for separately contacting the spin-resolved edge states (see Fig.~\ref{fig1}c). c) Temperature dependence of $I_{\uparrow}$ (red) and $I_{\downarrow}$ (blue) currents shows enhancement of relaxation between SRESs with increasing temperature. Thermally mediated mixing of currents becomes negligible at $T=250~{\rm mK}$.\label{fig2}}
 \end{center}
\end{figure}

We first measured the two-terminal magneto-current at $T=250$ mK in order to locate the plateau associated with a
number of filled LLs in the bulk $\nu$ equal to $2$
(see Fig.~\ref{fig2}a). The working point was set in the center of the plateau, i.e. at $B = 4.75$ T.
The two SRESs can be separately contacted as schematically shown in Fig.~\ref{fig1}c by negatively biasing the gates G1 and G2 at a voltage $V_G^*$, such that the filling factor below the corresponding top gates becomes $\nu=1$ and one edge channel only is allowed underneath the gates.
The actual $V_G^*$ value can be determined by measuring the currents $I_{\uparrow}$ and $I_{\downarrow}$ as a function of $V_G$ (see Fig.~\ref{fig2}b). 
When inter-edge coupling is suppressed by applying $V_a = -3~{\rm V}$ to all the nanofingers, we find that spin up electrons are entirely transmitted (yielding a current $I_{\uparrow}$ of about 1 nA, as expected for a single channel of unit quantized resistance $h/e^2 \approx 25.8~{\rm K}\Omega$), while the spin down current $I_{\downarrow}$ is nearly zero for $V_G^*=-0.47\pm0.08~{\rm V}$ (see Fig.~\ref{fig2}b).
In agreement with \cite{MullerPRB1992}, 
this implies the absence of significant spin flip processes over the distance of about $100~\mu{\rm m}$ traveled by the co-propagating SRESs when the magnetic fingers are deactivated.
\begin{figure}[t]
 \begin{center}
  \includegraphics[width=6 cm]{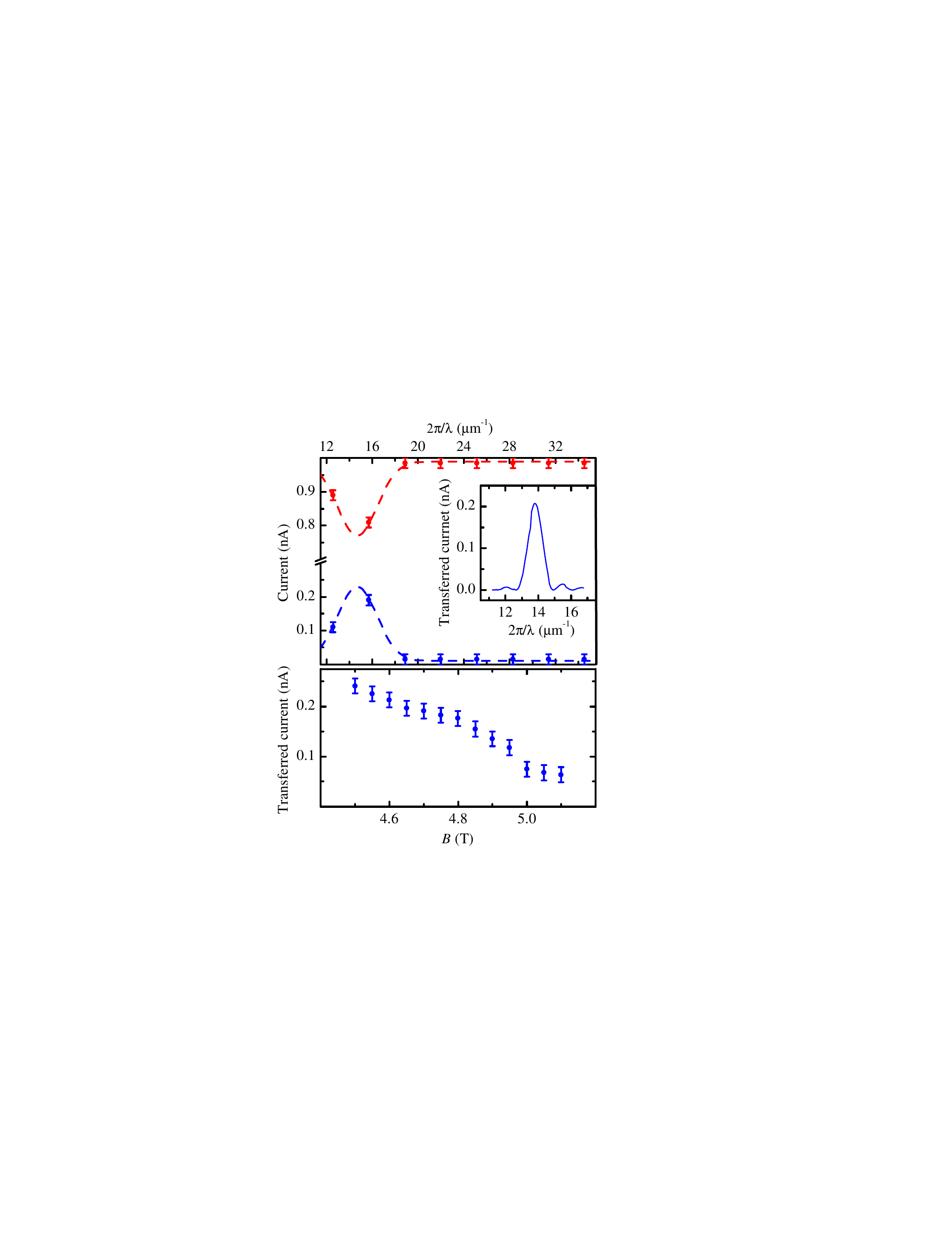}
    \caption{(Color online) Upper panel: Plot of the transmitted currents $I_{\uparrow}$ (red) and transferred current $I_{\downarrow}$ (blue) as a function of the inverse periodicity of the activated (by applying $V_a = 0$) set of nano-fingers at the working point $B=4.75$ T and $T=250$ mK.
The measured current $I_{\uparrow}$ and $I_{\downarrow}$ are guided by the dashed line which demonstrates selectivity of nano-magnet at periodicity between  $\lambda=400~{\rm nm}$ and  $500~{\rm nm}$. The inset shows a numerical simulation of transferred current  which in the absence of the 
static disorder and/or inelastic mechanisms predicts a width of the peak that scales inversely on $\Delta X$ as in Eq.~(\ref{Eq:t12amp}).
Lower Panel: measured transferred current $I_{\downarrow}$ as a function of the perpendicular magnetic field $B$ for the nano-magnet array of periodicity $\lambda=400~{\rm nm}$.
 \label{fig3}}
 \end{center}
\end{figure}
For completeness, Fig.~\ref{fig2}c shows the dependence of the currents $I_{\uparrow}$ and $I_{\downarrow}$ on temperature: SRESs fully relax only for $T\sim 1.6~{\rm K}$ (i.e. $1/(k_B{\rm T}) \approx 7.2~{\rm meV}^{-1}$), while edge mixing becomes negligible at our working point $T = 250$ mK. Moreover, analyzing our data as in Refs.~\cite{MullerPRB1992} we can conclude that the relaxation length is of the order of 1 cm at $T=250$~mK.

The upper panel of Fig.~\ref{fig3} shows the measured $I_{\uparrow}$ and $I_{\downarrow}$  when coupling occurs at several different individual arrays (one at a time) as identified by their $2\pi /\lambda$ value.
Since inter-edge coupling leads to charge transfer between the two spin-resolved edge channels it results in
a decrease of $I_{\uparrow}$, with the consequent increase of $I_{\downarrow}$ while the total current remains constant at about $1~{\rm nA}$. Note that 
current transfer is significant only for a specific interval of $\lambda$ values: indeed a 
 resonance peak appears  to occur at $\lambda_{res}$ between $400$ and $500~{\rm nm}$.
Such behavior is consistent with Eq.~(\ref{Eq:t12amp}) and with  a more refined theoretical analysis  based on the Landauer-B\"uttiker transport formalism~\cite{LAND} which
 we have solved numerically in order to go beyond the result of first-order perturbation theory ~\cite{KNIT} (see inset of the upper panel of Fig.~\ref{fig3} and SM). 
Static disorder and/or inelastic mechanisms induced, e.g.  by the finite temperature and Coulomb interactions,
may affect the resonance, 
resulting in a broadening of the 
current peak versus $2\pi/ \lambda$. 
 Importantly, if the fingers were an incoherent series of scatterers one should expect a monotonic $\lambda$-dependence of the charge transfer~\cite{DAVIDE}, while the observed non-monotonic selective behavior of the current suggests an underlying constructive interference effect. 

For the case of $\lambda=400~{\rm nm}$, the lower panel of Fig.~\ref{fig3} shows the dependence of transferred current $I_{\downarrow}$ on the perpendicular magnetic field $B$ when the latter spans the $\nu=2$ plateau (see Fig.~\ref{fig2}a).
The monotonic decrease of $I_{\downarrow}$ is a consequence of at least three 
combined effects: (i) the ratio $|\vec{B}_\||/B$ decreases as $B$ is increased, so that the net effect of the in-plane magnetic modulation is weakened; (ii) the magnetic length decreases with increasing $B$, causing the reduction of the spatial overlap of the transverse wavefunctions; (iii) the change of SRES spatial configuration with increasing magnetic field due to interaction effects~\cite{SGTIP,DEMP}.

\begin{figure}[t]
 \begin{center}
  \includegraphics[width=7 cm]{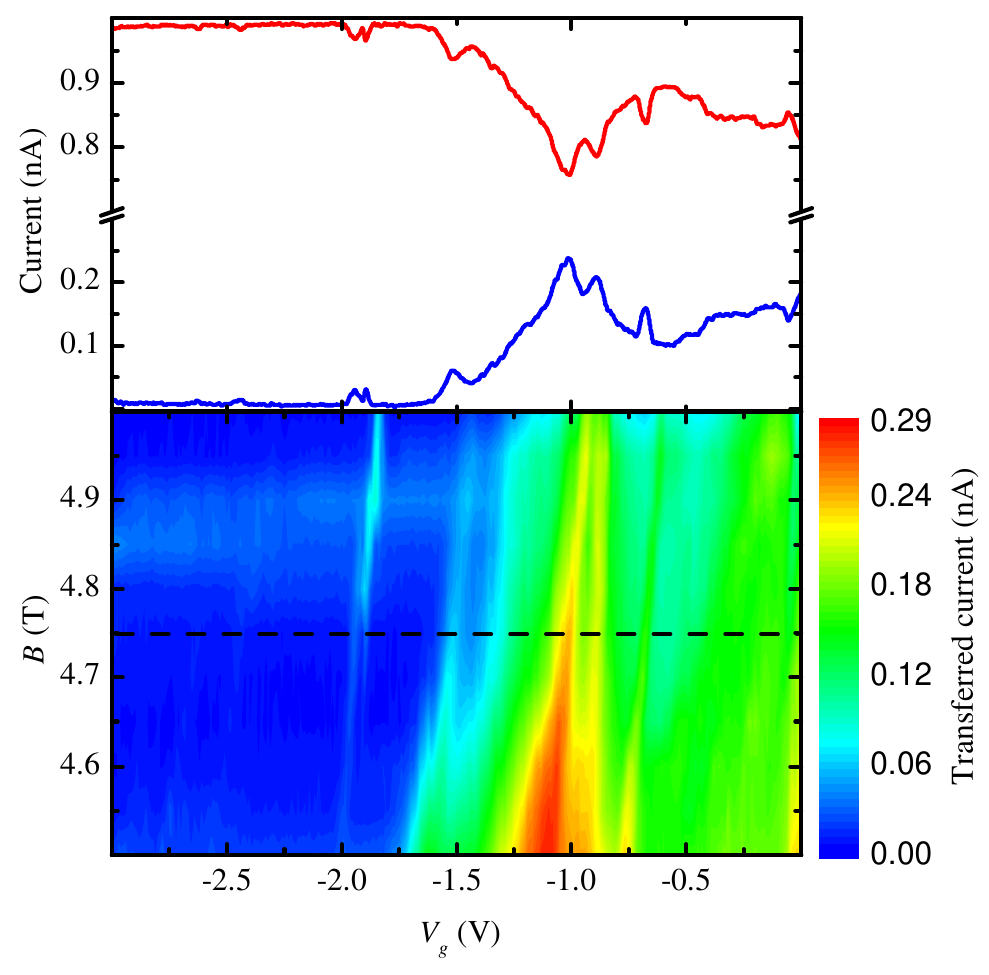}
    \caption{(Color online) Upper panel: Dependence of the transmitted current $I_{\uparrow}$ (red) and transferred current $I_{\downarrow}$ (blue) upon the voltage $V_a$ applied to activated nano-finger of periodicity $\lambda=400~{\rm nm}$ at $B=4.75~{\rm T}$.
      For $V_a<-2.0~{\rm V}$ the fingers are effectively decoupled from the SRESs with negligible transfer current; for $V_a\simeq 0$ instead the
    edges feel the presence of the fingers and a non-zero transfer of current is evident. For intermediate values of $V_a$ a series of pronounced peaks in $I_{\downarrow}$ are evident.
     Lower panel: contour plot of  $I_{\downarrow}$
    upon $V_a$  and $B$ for the nano-finger set of periodicity $\lambda=400~{\rm nm}$. The horizontal line
    indicates the center of the $\nu=2$ plateau ($B=4.75~{\rm T}$).  }\label{fig4}
 \end{center}
\end{figure}

Apart from activating/deactivating the various nano-finger sets, the voltage $V_a$ can also be used as an external control to adjust the resonant mixing condition.
Figure~\ref{fig4} shows the measured transferred current $I_{\downarrow}$
as a function of $V_a$ and $B$ for the array of periodicity $\lambda=400~{\rm nm}$ (similar data were obtained for different $\lambda $, see SM). The pronounced features present for intermediate values of $V_a$ show that the coupling between SRESs can be controlled and amplified. Remarkably, a charge transfer of $28\pm 1 \%$ was achieved at $B = 4.5$T with  $V_a \approx -1.1~{\rm V}$. At large negative $V_a$'s the SRESs are pushed away from the region where the magnetic fringe field is present and, as expected, the coupling vanishes.
The same Fig.~\ref{fig4} reveals additional resonances occurring at specific values of $V_a$. A non-monotonic dependence of the local value of $\Delta k$ on $V_a$,
  can be invoked to explain these features. {A system simulation shows that a local change of the confinement potential in the proximity of the associated nano-fingers modifies the relative distance of the 
SRESs  and hence the local value of $\Delta k$ in a non-monotonic way  (see SM).
More precisely, for low $V_a$ the fingers act as top gates for the underlying edge states: the transverse distance between SRESs 
 can locally increase and reach a maximum as $V_a$ gets negative, since  $\Psi_{\downarrow}$ and $\Psi_{\uparrow}$ 
 are pushed away from the finger region, one after the other. As we further increase $V_a$ the transverse distance between the SRESs  increases again. It is worth stressing, however, that the process just described is not necessarily smooth: electron-electron interaction may in fact induce abrupt transitions in SRESs distances when the slope of the effective local potential decreases below a certain critical value which depends on the details of the sample properties~\cite{DEMP} (also the
  gate voltage can influence the Fermi velocity, as shown in edge magnetoplasmons time-of-flight experiments~\cite{EMP}). The trajectories of SRESs are unknown and (differently from what shown in the graphical rendering of Fig.~\ref{fig1}a)
   are likely to be outside the regions corresponding to the projections of the fingers when a significant voltage is applied. Nevertheless non-linear repulsive effect is expected to be effectively active in the experiment where the electrostatic potential profiles extends much beyond the length of the fingers.
Moreover, the functional dependence of the potential induced by $V_a$ upon the longitudinal coordinate $x$ presents also an oscillatory behavior with periodicity $\lambda$. As a consequence of the adiabatic evolution of the edges, their transverse distance will also show such oscillations. A detailed modeling of the observed resonance features would require to take fully into account these effects and is beyond the scope of the present paper. However it clearly deserves further investigation as it represents a positive feature of the system, since any value of the modulation periodicity $\lambda$ has typically more than one value of $V_a$ that can fulfill the resonant condition.

Our proposal provides a way to realize
beam splitters for flying qubit using topologically protected SRESs.
 It employs a nanofabricated periodic magnetic field operated at a resonant condition 
  which enhances quite significantly the weak magnetic field produced by the Cobalt nanomagnets.
Already at $T =250$ mK the effect is significant and should be enhanced at lower temperatures.

This work was supported by MIUR through FIRB-IDEAS Project No. RBID08B3FM and by EU through Projects SOLID and NANOCTM. 
We acknowledge useful discussions with N. Paradiso and S. Heun.

\newpage

\begin{widetext}

\section*{Supplemental Material}

Here we describe the theoretical approach to numerically simulate the transport properties of the proposed device. On this basis we
discuss a mechanisms that, for a given periodicity of the fingers, produces multiple resonances in the transferred current as
we vary the gate voltage $V_a$. We also report a comparison between the transferred current computed at 
first order and the exact numerical solution, showing that indeed the former is able to detect the resonant condition and we provide an estimate of the resonant periodicity of the array.  Finally we present some extra data of the measured transferred current as a function of $V_a$ and $B$ for different periodicities $\lambda$ of the fingers.

\subsubsection*{Numerical Simulations} 
We performed numerical simulations by modeling the device through a tight-binding Hamiltonian
describing a Hall Bar about 100 nm wide (hard wall confinement potential), 
with lattice spacing $a=2.8~{\rm nm}$~\cite{key-1}.
The magnetic field $B=4.75~{\rm T}$, corresponding to a cyclotron gap $\hbar\omega_c\approx 7.85~{\rm meV}$, is introduced by Peierls phase factors on the hopping amplitudes.
The Zeeman gap has been taken to be $E_{\rm Z}=\hbar\omega_c/17$,
corresponding to an effective value of  $g^* = 1.4$ (as shown below this value leads to a $\lambda_{res}$ consistent with the experimental value reported in Fig.3 of the main text). This value, 
which is larger than the usual value for GaAs (i.e. $|g^*|=0.44$), compensates the reduced distance between the edge sates arising from the use of an hard wall approximation in our 
simulations. Notice that this value is compatible with the estimation of $\lambda_{\rm res}$ one gets by using the  following simple heuristic  analysis.
Indeed assuming adiabatic following of the confinement potential by the Landau levels, the transverse spatial separation $\Delta Y$ between the SRESs can be evaluated by
an energy-balance argument: the external confinement field must work
against the energy gap $\Delta \epsilon$ in order to make the channels degenerate in
energy. For a linear confinement potential $V_c(y)=e Ey$, this yields
\begin{equation}\label{eq:DeltaXexperimentalEstimation}
eE\Delta Y=\Delta\epsilon\;. 
\end{equation}
While it is experimentally very tricky to measure the edge separation
for SRESs, several solid hints can be extrapolated for the separation $\Delta Y_{c}$
of cyclotron-resolved edge states~\cite{Muller}. More specifically, we could use the
experimental value of $\Delta Y_{c}$ for $\Delta\epsilon=\hbar\omega_{c}$,
to infer by proportionality the $\Delta Y$ for spin-resolved channels:
$\Delta Y\simeq\frac{\epsilon_{z}}{\hbar\omega_{c}}\Delta Y_{c}$.
By employing the measurements of Ref. \cite{Nicola} performed at $\nu=4$
in similar experimental conditions, we are lead to $\lambda_{\rm res} \simeq360~{\rm nm}$ which, 
considering the different approximations adopted in the two approaches (e.g.  linear confinement vs. 
sharp potential), is in good agreement with the number we obtained from the simulations.
The above analysis give us also the opportunity  of stressing that 
the mixing of spin-degenerate, energy-degenerate edge 
channels belonging to different LLs would require a spatial modulation of the perturbative field with periodicity on order of few 
\AA$\,$ that  is practically impossible to engineer. In the absence of perturbation potentials, this yields a wave-vector difference of the two spin resolved edges of the order of  $\Delta k \simeq 13.8$~{\rm nm}$^{-1}$ corresponding to a value of the resonant condition
 $\lambda_{\rm res}= 2\pi/\Delta k \simeq 450$ nm and to  an overlap integral of the wavefunctions 
$\gamma = \int dy \chi_{0,k_{\uparrow}}(y)\chi_{0,k_{\downarrow}}(y)$  of $\simeq 0.97$  (number obtained by using the wave-functions $\chi_{0,k_{s}}(y)$ for hard wall confinement~\cite{MAC}). 

 Transport properties are computed through the recursive-Green function
algorithm of the KNIT Numerical Package~\cite{knit}. 

A nano-magnet finger array is placed at the right boundary of the 2DEG (short dashed rectangles in Fig.~\ref{figSM1}b) and produces a periodic magnetic field $\vec{B}_{\parallel}$ in the $x$-direction, which extends on the Hall bar in the $y$-direction for 40 nm. 
The field $\vec{B}_{\parallel}$ induced on the 2DEG by a single rectangular magnetic finger, which can be calculated exactly, is shown through a vectorial plot in Fig.~\ref{figSM1}a (its maximum amplitude reaching values of the order of 
50mT~\cite{Cocriticalfield}).
Selectivity of edge channels is realized by placing top gates in the injection and detection electrodes (long dashed rectangles in Fig.~\ref{figSM1}b) which induce an electrostatic potential so to set a filling factor equal to 1 underneath the gates.
As a result, the outer edge state only is allowed to enter the electrode.
This is clearly demonstrated by the charge density plots of the inner ($\Psi_{\downarrow}$) and the outer ($\Psi_{\uparrow}$)
SRES shown in the upper and lower panels, respectively, of Fig.~\ref{figSM1}c.

\begin{figure}

\begin{centering}
\includegraphics[width=10 cm]{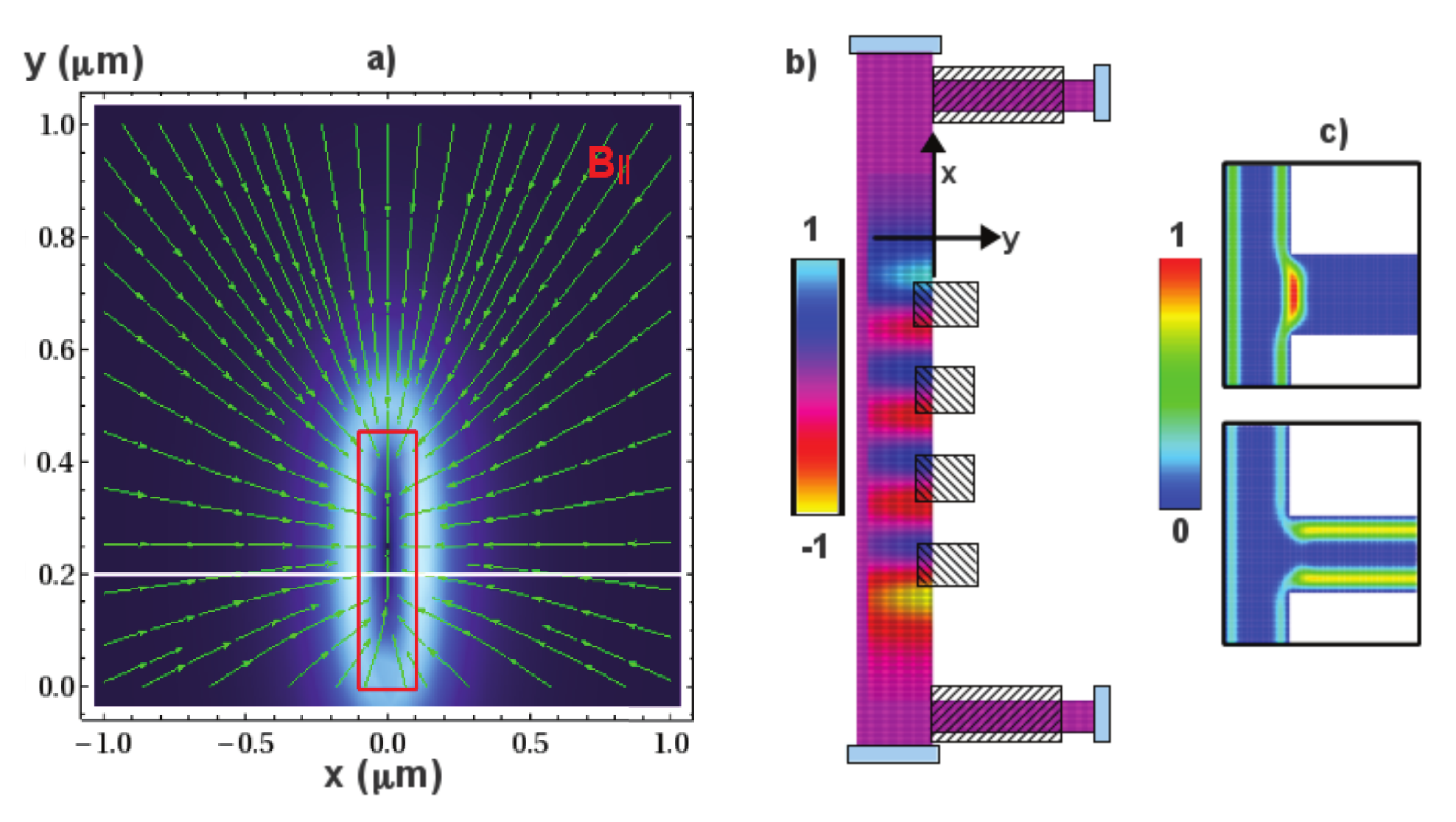}
\par\end{centering}

\caption{ 
a) vectorial field plot and density plot of the in-plane component $\vec{B}_{\parallel}$ of the magnetic field computed for a single magnetic finger (indicated by the red rectangle): 
the direction of the field in the plane is represented by the green arrows while its intensity is represented by the different colors of the background  
(lighter hue corresponding to higher intensities). The white line in the plot indicates where the physical edge of the mesa.
b) Schematic of a simulated Hall bar device 
where the spatial distribution of the $x$-component of  the inhomogeneous magnetic field $\vec{B}_\|$ 
 generated by a finger array consisting of just four fingers (short dashed rectangles in the figure) is highlighted;
the top gates used for the selective injection and detection electrodes are represented by long dashed rectangles.
c) Charge density plots of the $\Psi_{\downarrow}$ 
(lower panel) and $\Psi_{\uparrow}$ 
(upper panel) SRES in the proximity of the simulated top gates that guarantee their selective population. \label{figSM1}}

\end{figure}

\subsubsection*{Multiple resonances} 

As discussed in the main text, when the negative value of the voltage $V_a$ applied to the fingers increases, the actual path followed by the edge channels is deformed so that their local separation, and hence their wave vector difference $\Delta k$, can vary in a non-linear fashion.
Using the method detailed in Ref.~\cite{key-3}, we numerically determine $\Delta k$ as a function of $V_a$ for a single long finger extending 19.7 nm in the $y$-direction.
The resulting resonant periodicity, defined as $\lambda_{\rm res}=2\pi/\Delta k$ and plotted in Fig.~\ref{figSM2} as a function of $V_a$ (blue line), first decreases, reaching a minimum, and thereafter slowly increases.
Such behavior reflects the fact that the two edge channels are progressively expelled from underneath the finger one after the other.
The process is pictorially described by the three cartoons a), b), c) on the right side of Fig.~\ref{figSM2}. Here the arrows describe the position of the two edges in the mesa while the dashed rectangle represents the region of the fingers: the configuration a) corresponds to the case in which $V_a$ nullifies (both edges lie below the fingers); configuration b)  corresponds
to the situation in which   the inner state only is expelled (when this happens $\lambda_{\rm res}$ reaches its minimum value); finally configuration c) corresponds to the case 
of very large negative value of $V_a$ when both edges are completely expelled from the region beneath the fingers: this effect is illustrated in the inset to  Fig.~\ref{figSM2}a where
a numerical  simulation of the charge  density  of the outer channel  $\Psi_{\uparrow}$
shows that the corresponding edge state has been pushed away from the finger region (in the plot the electrostatic repulsion has been taken homogeneous).

\begin{figure}[t]
\begin{centering}
\includegraphics[width= 11  cm]{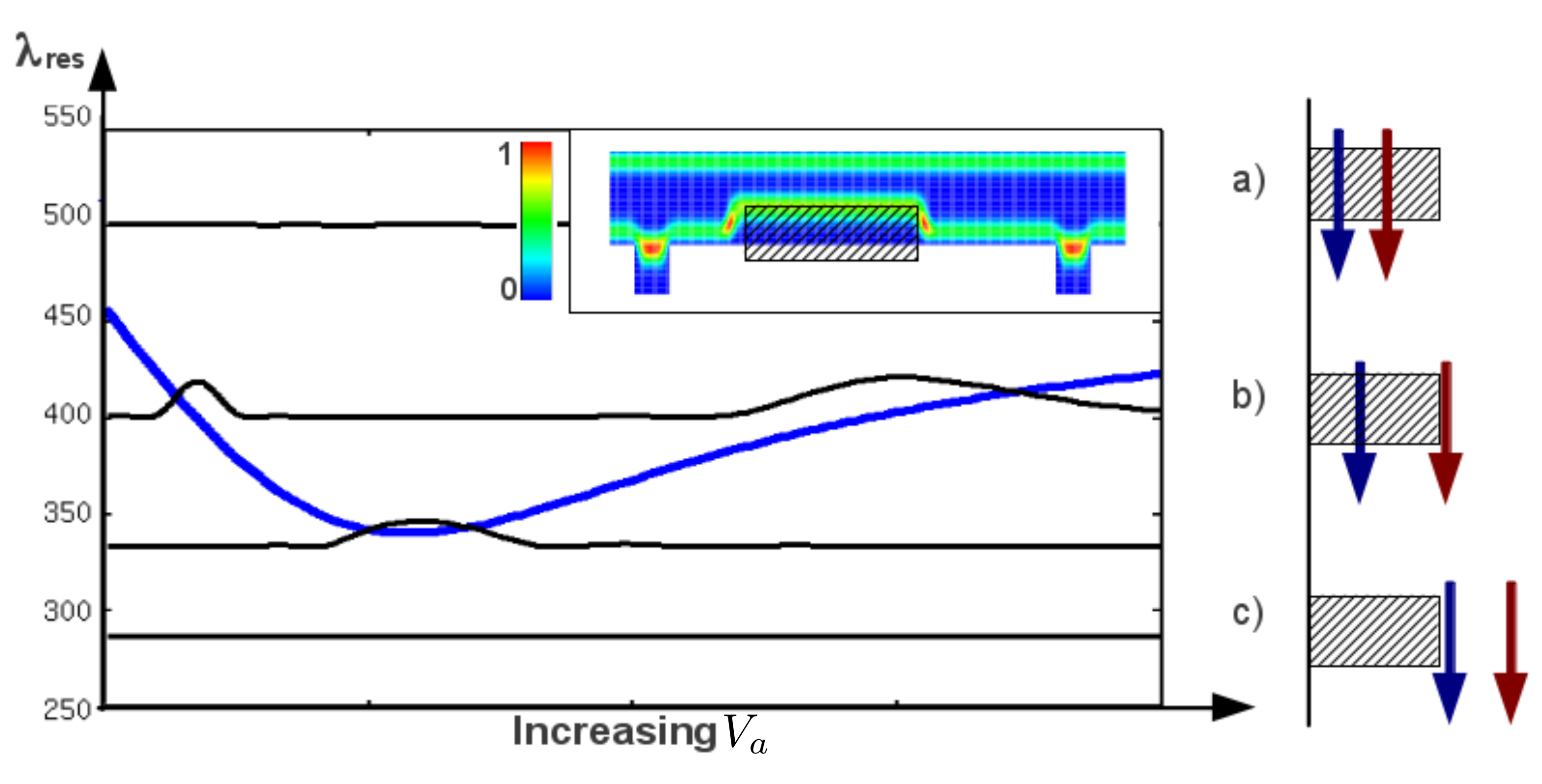}
\par\end{centering}
\caption{Left panel: The blue curve indicates the resonant period as function of the
finger voltage $\lambda_{\rm res}\left(V_a\right)$, while the different black
curves represent the transferred currents $I_{\downarrow}(V_a)$
for simulated devices consisting of  (from top to bottom) 12 magnetic fingers spaced 500 nm, 15 fingers spaced 400 nm, 18 fingers spaced 333 nm and 21 fingers spaced 286 nm.
For the sake of clarity the scales of the current for each curve have been set in arbitrary unit and 
their zero levels have been shifted to match with the periodicity of the associated $\lambda$.
 Inset: example of the simulated Hall-bar, where the charge density solution for the outer edge channel is plotted for large negative $V_a$ (here 
 the electrostatic repulsion is considered homogeneous in the finger region).
Right panel: pictorial view of the repulsion effect of the edges (represented in the picture by the arrows)
 at the origin of the non-monotonic behavior of the resonance condition with the increase of $V_a$ (shaded rectangles represents the area underneath the fingers); a) $V_a=0$, b) maximum separation corresponding to the minimum of the resonance curve in the right panel, c) total repulsion of the edge channels at large values of $V_a$.\label{figSM2}}
\end{figure}

In the left panel of  Fig.~\ref{figSM2} we report also the  value of the computed  transferred current $I_{\downarrow}$
for arrays of different periodicities $\lambda$ as a function of $V_a$ (black curves of the figure --- see caption for details). 
We notice that in correspondence of the matching between $\lambda(V_a)$ (blue curve in picture)
with the periodicity $\lambda$ of the finger, $I_{\downarrow}$ shows a peak (otherwise it is zero). This indicates that 
the resonant condition discussed  in the main text, can be met for more than one value of $V_a$ depending on the periodicity. Such an effect is
in qualitative agreement with Fig.~4 of the main text (see also the last section of the Supplementary Material) which, for fixed $\lambda$, shows the presence of resonant peaks
for intermediate values of $V_a$.

\subsubsection*{Comparison between first order and exact numerical solution} 
In the main text we used  first order perturbative  analysis to show the existence of a resonant condition for the finger periodicity (i.e. $\lambda_{\rm res}= 2\pi/\Delta k$).
Here such result is compared with the exact solution obtained by computing the transferred current via the tight-binding  recursive Green's function method~\cite{key-1,knit}
detailed in the first section of the supplementary material.
 
The resulting plots are shown in Fig.~\ref{fig5}: 
we chose the cyclotron gap $\hbar\omega_c\approx 7.85~{\rm meV}$, corresponding to a magnetic field $B=4.75~{\rm T}$, and Zeeman gap $E_{\rm Z}=\hbar\omega_c/17$, such to produce the resonant peak between $2\pi/0.4$ and $2\pi/0.5~\mu{\rm m}^{-1}$. The longitudinal extent of the nano-magnet array is $\Delta X=6~\mu{\rm m}$ and the in-plane component of the associated magnetic field are calculated exactly. The analytic expression of the transferred current, following the perturbative approximation detailed in the main text, has been evaluated for $\Delta k\approx 13.8~\mu{\rm m}^{-1}$ and properly rescaled to match the result of the simulation. The good agreement between the two curves shows that the perturbative approach is sufficient to capture the resonant behavior of the device.

\begin{figure}[t]
 \begin{center}
  \includegraphics[width=6cm]{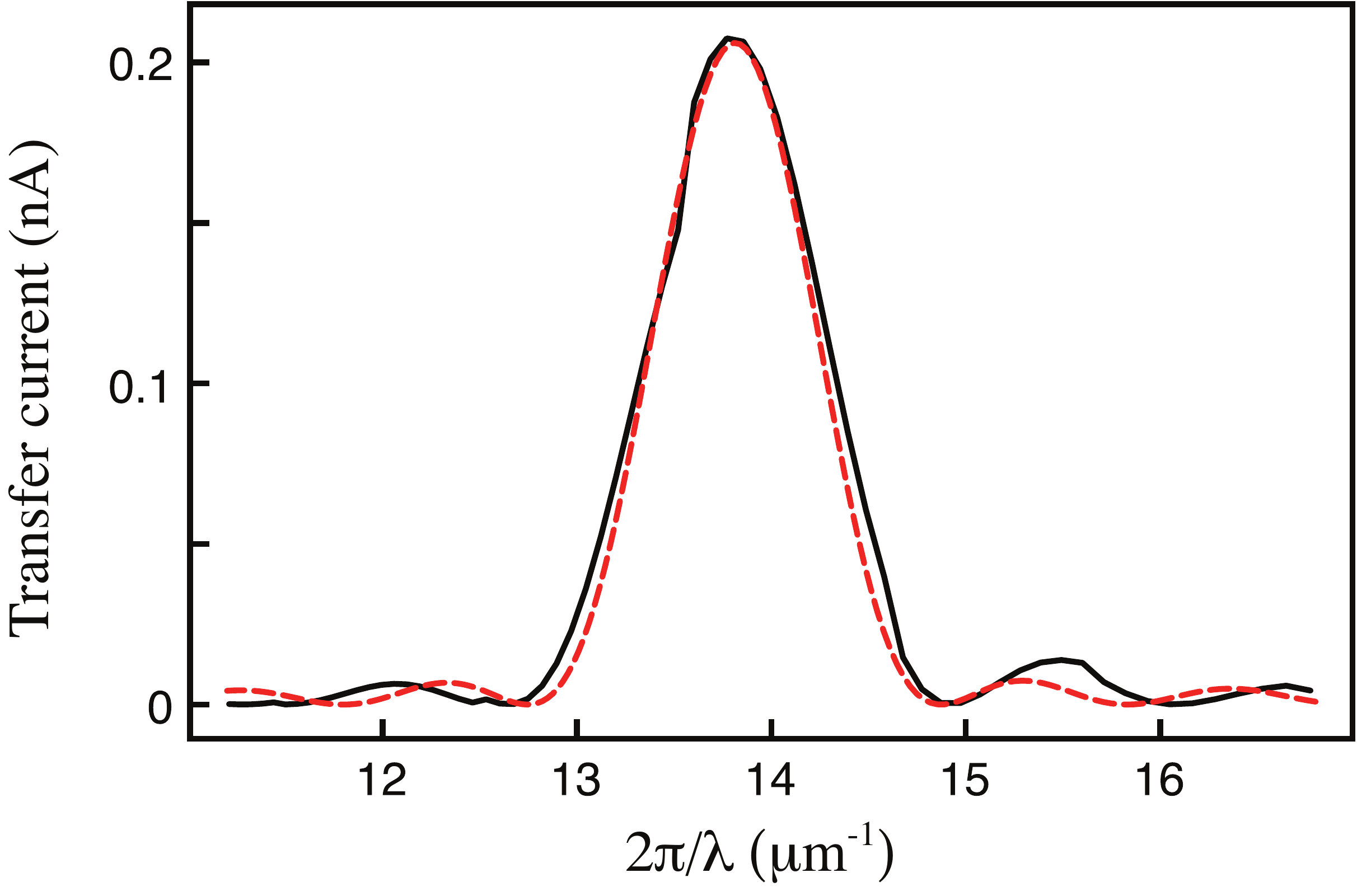}
    \caption{Comparison between the transfer current $I_{\downarrow}$ obtained from the perturbative approach Eq.~(1) of the main text (red dashed curve) and from a  numerical calculation (black solid curve). The red dashed curve is rescaled in order to match the maximum value obtained with the numerical calculation. 
    \label{fig5}}
 \end{center}
\end{figure}

\subsubsection*{Transferred current as a function of magnetic field and gate voltage} 
The reproducibility of controlled coupling of spin-resolved edge states at filling factor $\nu = 2$ is shown in Fig.~\ref{figSM4}. The measurements are performed under the same procedure and experimental conditions described in the main paper and several weeks after the experiments reported in the main paper. The data for the array with $\lambda = 400$nm refer to a different cool down of our sample and a slightly different electron density. However, the generic features of the measured transferred current $I_\downarrow $ are similar to the color plot reported in the main paper. We emphasize that at zero bias voltage ($V_a $ = 0), the transferred current $I_\downarrow $ for the namomagnet array of periodicity $\lambda = 400$ nm  is maximum compared to the other nanomagnet arrays and decays with increasing applied perpendicular magnetic field B, spanned over the $\nu $ = 2 QH plateau. Therefore, the result is consistent with our theoretical understanding and the reported experimental results in Figs. 3 and 4 of the main paper. Moreover, at intermediate values of $V_a $ from 0 to -2.25 V, several resonant peaks in the transferred current $I_\downarrow $ for both the nanomagnet arrays of periodicities of $\lambda = 400$ nm  and $333$ nm  appear and shift quasi linearly towards higher values of $V_a $ with increasing perpendicular magnetic field B. The transferred current $I_\downarrow$ becomes significantly low as the artificial coupling induced by the nanomagnet arrays vanishes for $V_a $ less than -2.25 V. It is also notable that the signal strength of the transferred current $I_\downarrow$ is lower than that reported in the experiments described in the main paper (Figs. 3 and 4). This signal reduction indicates oxidation of Cobalt nanomagnet fingers leading to a reduction of the parallel fringing field $\vec{B}_\|$ due to formation of antiferromagnetic CoO  \cite{Cocriticalfield}.

\begin{figure}
\begin{centering}
\includegraphics[width=8cm]{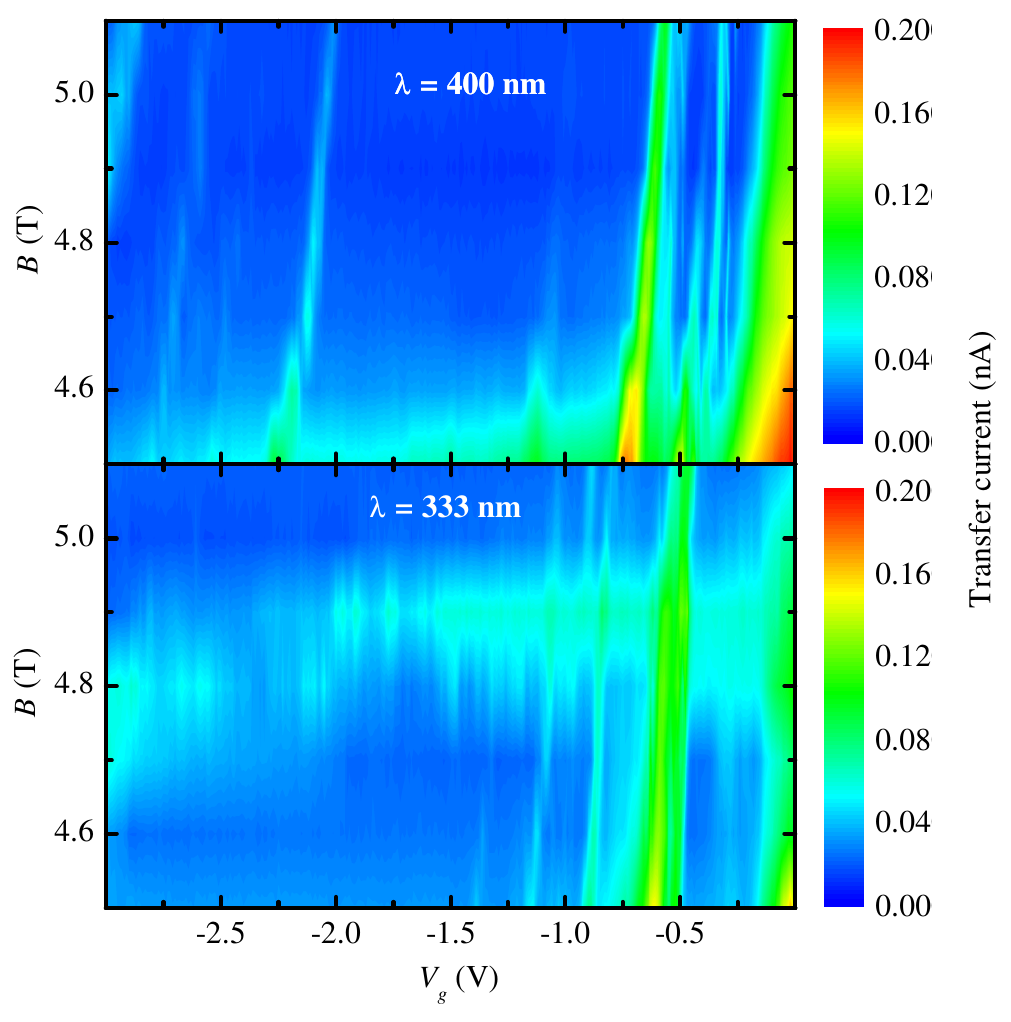}
\par\end{centering}
\caption{Color plot of measured transferred current $I_\downarrow$ in the magnetic field (B) -  gate voltage ($V_a$) plane for nanomagnet arrays of periodicities $\lambda = 400$nm and $\lambda = 333$nm. The measurement is performed on the same sample after a different cool down.\label{figSM4}}
\end{figure}

\end{widetext}

\end{document}